\documentclass[sigconf,review=False]{acmart}
\usepackage{listings}
\AtBeginDocument{%
  \providecommand\BibTeX{{%
    \normalfont B\kern-0.5em{\scshape i\kern-0.25em b}\kern-0.8em\TeX}}}

\setcopyright{acmcopyright}
\copyrightyear{2026}
\acmYear{2026}
\acmDOI{XXXXXXX.XXXXXXX}

\acmConference[SIGCSE Virtual '26]
{the 2nd ACM Virtual Global Computing Education Conference}
{November 12--15, 2026}
{Virtual Event}

\acmISBN{978-1-4503-XXXX-X/18/06}

\begin{document}


\title{Exploring the Value of Diverse LLM Explanations in Introductory Programming 
} 

\author{Seth Bernstein}
\affiliation{
  \institution{Temple University}
  \city{Philadelphia}
  \state{PA}
  \country{United States}}
\email{sethbern@umich.edu}
\orcid{0000-0002-7552-5448}

\author{Paul Denny}
\affiliation{%
  \institution{University of Auckland}
  \city{Auckland}
  \country{New Zealand}}
\email{paul@cs.auckland.ac.nz}
\orcid{0000-0002-5150-9806}

\author{Juho Leinonen}
\affiliation{%
  \institution{Aalto University}
  \city{Espoo}
  \country{Finland}}
\email{juho.2.leinonen@aalto.fi}
\orcid{0000-0001-6829-9449}

\author{Kush Patel}
\affiliation{%
  \institution{Temple University}
  \city{Philadelphia}
  \state{PA}
  \country{US}}
\email{kushrp@temple.edu}
\orcid{0009-0008-0356-3294}

\author{Rayhona Nasimova}
\affiliation{%
  \institution{Temple University}
  \city{Philadelphia}
  \state{PA}
  \country{US}}
\email{rayhana.nasimova@temple.edu}
\orcid{0009-0005-8113-1858}

\author{Matt Littlefield}
\affiliation{%
  \institution{Temple University}
  \city{Philadelphia}
  \state{PA}
  \country{US}}
\email{matt.littlefield@temple.edu}
\orcid{0009-0008-1614-1875}

\author{Stephen MacNeil}
\affiliation{%
  \institution{Temple University}
  \city{Philadelphia}
  \state{PA}
  \country{US}}
\email{stephen.macneil@temple.edu}
\orcid{0000-0003-2781-6619}

\renewcommand{\shortauthors}{Bernstein, et al.}

\begin{abstract}
Large Language Models (LLMs) have shown the potential to generate code explanations that surpass those of peers in quality, offering promising opportunities for computer science education. While these explanations may not yet match the depth and clarity of instructor-provided explanations, research in computational creativity highlights that the quantity and diversity of ideas can often outweigh a singular focus on quality. Inspired by this, we explore whether combining multiple diverse explanations, each emphasizing distinct aspects (e.g., function, concept, goal), can enhance students’ understanding of programming exercises compared to generic explanations that do not emphasize distinct conceptual aspects. In our study  971 first-year computing students were randomly assigned either diverse or generic LLM-generated explanations for two programming exercises. Students completed multiple-choice and open-ended questions for each exercise, followed by Likert-scale questions and open-ended reflections. Our findings outline patterns in student performance and perceived cognitive load across the two explanation conditions. These findings highlight how variation in explanation emphasis may relate to learner engagement and understanding. Across participants, open-ended response accuracy was consistently about 7.7\% higher when students received diverse explanations, with no difference in perceived cognitive load.
\end{abstract}





\maketitle

\section{Introduction}

Teaching students code comprehension skills has been a long-standing and important pedagogical goal within computing education~\cite{corney_pt,lister2004multi}. While textbooks and online tutorials offer explanatory materials, these static resources lack the flexibility to adapt to the diverse needs of individual students~\cite{lahtinen2005study}. Recent advances in Large Language Models (LLMs) enable the immediate, automatic generation of diverse code explanations~\cite{openai_codex, Jury2024worked, raihan2025large}. While prior work has shown that LLMs can generate diverse explanations~\cite{macneil2023experiences, leinonen2023comparing}, it remains unclear how presenting multiple explanations with intentionally distinct semantic emphases together affects novice programmers’ understanding.

Insights from other fields, such as computational creativity, suggest that diverse ideas may be more beneficial than relying solely on a single, high-quality option~\cite{siangliulue_toward_2015}. This aligns with \textit{Variation Theory}~\cite{lo_towards_2012}, which holds that learners grasp a concept when they see systematic variation that exposes its critical features, helping them distinguish it from related ideas. Motivated by this, we investigate whether incorporating multiple LLM-generated \textbf{diverse code explanations}, each emphasizing distinct aspects (e.g., function, concept, goal), can enhance students’ understanding of programming exercises compared to relying on generic explanations. These were compared with \textbf{generic explanations}, which cover the snippet in an all-purpose way, rather than highlighting a specific aspect.

To address this gap, we conducted a large-scale, between-subjects study with \textbf{971 first-year computing students}. Participants were randomly assigned to receive either \textit{three diverse LLM-generated explanations} or \textit{three generic LLM-generated explanations} when learning to solve two programming exercises. We collected answers to multiple-choice questions (MCQ) and open-ended questions (OE) to gauge understanding, and we used Likert ratings with reflections to assess helpfulness, clarity, and cognitive load, as good explanations should boost learning without taxing limited working memory~\cite{stadler2024cognitive}.

\begin{itemize}
    \item[\textbf{RQ1}] Do diverse LLM-generated explanations improve students’ understanding of programming exercises more effectively than generic explanations?
    \item[\textbf{RQ2}] Do students perceive diverse explanations as more helpful or informative than generic explanations, and how does this perception align with their actual learning outcomes?
\end{itemize}

\section{Related Work}

Explanations play an important role in computing education. When students encounter unfamiliar or complex code, they often depend on explanations to make sense of syntax, semantics, and underlying logic~\cite{corney_pt, marwan2019}. Explanations can reduce confusion, support mental model development, and improve learning outcomes~\cite{venables2009closer}. As LLMs are increasingly used to generate explanations at scale~\cite{macneil2023experiences, leinonen2023comparing}, recent work has begun to examine how these AI-generated responses compare to student-written ones \cite{leinonen2023comparing}, and test whether models can craft programming diverse explanations such as for worked examples and based on analogies~\cite{openai_codex, bernstein_like_2024, Jury2024worked, prather2023robots}. Understanding how students engage with these explanations is key to determining their effectiveness and potential harms~\cite{bernstein_harms_2025}. 

\subsection{What Makes a Good Code Explanation?}

Explanations are only as useful as the learner's ability to interpret and understand the explanation. Clear and concise language can reduce cognitive load and help learners focus on the main ideas ~\cite{mayer_nine_2003,ainsworth_functions_1999}. Effective explanations also build on the learner’s background knowledge without introducing many unfamiliar concepts or ideas~\cite{mayer2005cognitive}. For instance, an explanation intended for a college freshman might focus on foundational concepts, whereas more advanced students could benefit from in-depth technical detail~\cite{lahtinen2005study}. The effectiveness of an explanation also depends heavily on the learner’s goal, whether they are trying to understand syntax, debug behavior, or generalize a concept ~\cite{corney_pt, sandoval_explanationdriven_2004}. Despite the value of personalized explanations, students often rely on static explanations in a textbook or crowdsourced explanations on Stack Overflow~\cite{dondio2020stack}. Additionally, students will adapt explanations to their own contexts~\cite{bettin2022semaphore}.

Prior work in artificial intelligence research and expert systems has identified several factors, including terminology, user sensitivity, abstraction, summarization, perspectives, linguistic competence, and feedback, that guide explanation clarity and utility \cite{bove_contextualization_2022, meaningful_user}. However, these studies primarily address single explanations rather than exploring how multiple, thematically distinct explanations may deepen conceptual learning in computing contexts.

\subsection{Diversity in Creativity and Learning}

Research in computational creativity has consistently shown the benefits of diverse exemplars to stimulate innovative thinking~\cite{dow_prototyping_2011, ward1999creative}. For instance, learners who encountered diverse solution types during instruction were better able to transfer concepts to novel problems~\cite{lo_towards_2012}. In a study of how humans use design inspiration, researchers found that ``diverse sets of examples from an idea map generate more diverse ideas than those seeing randomly selected examples''~\cite{cai_designaid_2023}. Similarly, in creative domains, uniform or homogeneous exemplars can lead to design fixation, whereas varied inputs support more flexible reasoning~\cite{siangliulue_toward_2015, cai_designaid_2023}.

By extension, having a broad set of explanations for the same snippet could enable novice programmers to see the code from multiple angles. For instance, one explanation might focus on core functionality, while another explores underlying concepts. Although maximizing diversity alone could dilute the quality \cite{siangliulue_toward_2015},  intentionally varying the semantic focus of the explanations while maintaining relevance may deepen comprehension, especially for first-year students learning fundamental programming tasks. 

\subsection{Using LLMs to Generate Code Explanations}

Recent studies have evaluated the capabilities of LLMs to create code explanations ~\cite{macneil2023experiences, macneil2022generating, leinonen2023comparing}. In one of the earliest evaluations of code explanations generated by LLMs, code explanations generated by Codex were correct 67.2 percent of the time~\cite{openai_codex}. Another study investigating the types of code explanations that can be generated demonstrated that GPT-3 could create explanations in diverse formats~\cite{macneil2022generating}. Explanations by LLMs have been shown to be rated as more useful by students than explanations written by their peers~\cite{leinonen2023comparing}. Prior work has also explored how LLMs can be used to personalize instruction by incorporating student interests and generating analogies relevant to their background~\cite{bernstein_like_2024, bernstein_analyzing_2024}.  However, while prior work has demonstrated that LLMs can generate and present multiple explanations in different formats or styles \cite{macneil2023experiences, macneil2022generating}, these studies did not explicitly investigate semantic diversity across explanations as a controlled instructional manipulation. In particular, \cite{macneil2023experiences} presented explanations in multiple formats but did not isolate conceptual dimension—function, concept, or goal—as the experimental variable, which is the focus of this study.

\section{Method}
\subsection{Study Context}
This study was conducted in fall 2024 during regular lab sessions for a first-year engineering programming course at a large public research university. The 12-week course, which uses C as the primary language, is required for all engineering students and is typically taken in their first year. There is an incentive for these students to perform well in the course as they choose their specialization based on their second-year GPA. A total of 971 students took part in the study. Students were not explicitly informed that the explanations they received originated from an LLM, and they could not ask follow-up questions about the source of these explanations. Of the 971 students, 785 provided sufficiently complete OE responses for analysis, while all 971 completed MCQ items. Ethics approval was granted by the institution (\textit{UAHPEC25279}).

\subsection{Materials}
\begin{figure}
\centering
  \includegraphics[width=\linewidth]{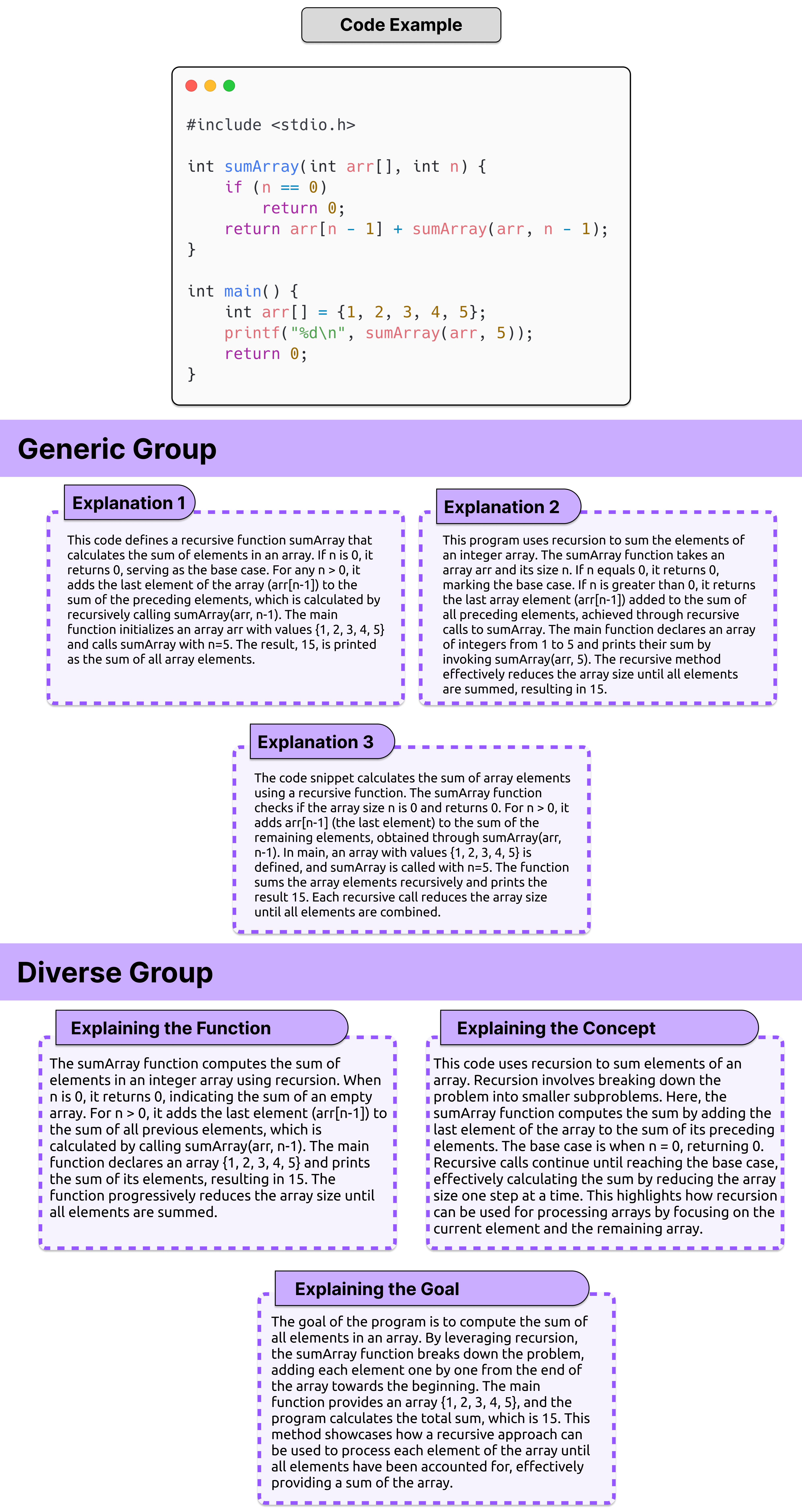}
  \caption{sumArray with code explanations (student view). Students saw the code example alongside three explanations from one condition: either the Generic group (top) or the Diverse group (bottom), never all six.}
  \label{fig:boat1}
\end{figure}

Three code snippets were prepared to enable counterbalancing, though each participant only saw two to limit cognitive burden during the lab session. For an example, see Figure~\ref{fig:boat1}. Additionally, we used GPT-4o  to generate explanations for each snippet. In the \textbf{Generic condition}, we issued the identical prompt, "Explain what this code does in plain text," three times and showed the three plain-text summaries together, ensuring visual parity with the diverse condition. For the \textbf{Diverse condition}, we issued three distinct prompts targeting function, concept, and goal as the three dimensions, generating one explanation per dimension. We selected function, concept, and goal as the three dimensions because prior work on code comprehension identifies these as distinct cognitive registers: understanding what code does, why the underlying mechanism works, and what the program is trying to accomplish~\cite{corney_pt,mayer_nine_2003}.

\subsection{Study Design and Procedure}
\begin{figure*}[htbp]
  \centering
  \includegraphics[width=\linewidth]{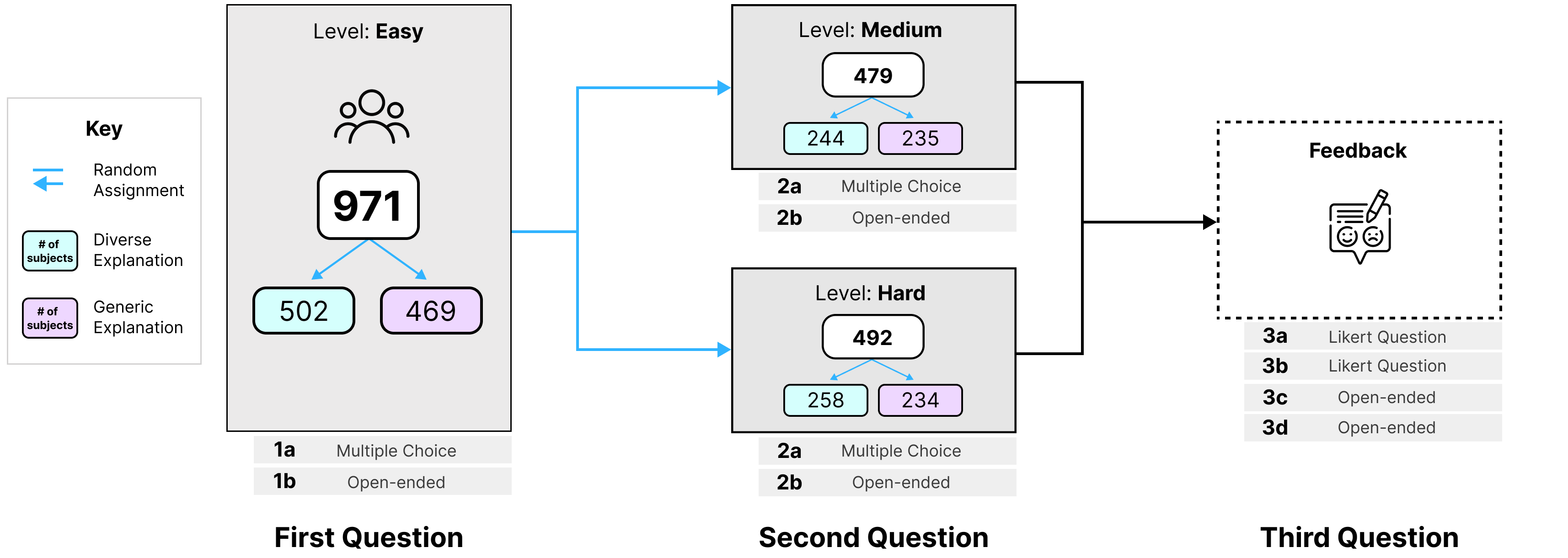} 

  \caption{Breakdown of participant flow through assessment. All students completed an easy problem before being randomly assigned to a second problem of medium or hard difficulty. Finally, they completed a feedback survey.}
  
  \label{fig:flow}
\end{figure*}

Participants were randomly assigned (between-subjects) in an online environment to receive either three \emph{Generic (G)} or three \emph{Diverse (D)} explanations for two programming snippets: \textit{sumArray} (Snippet 1) followed by either \textit{randomizeString} or \textit{countChar} (Snippet 2). This yielded four explanation sequences: G/G, G/D, D/G, D/D. This design was chosen to allow observation of order effects, specifically whether experiencing one explanation type first influenced perception of the second, without requiring a fully within-subjects design that would introduce fatigue across multiple lab tasks. Because each student viewed two snippets, the four explanation sequences produced eight analysis groups (one per sequence–snippet pair). After viewing the explanations for each snippet, students answered one multiple-choice and one open-ended question targeting core recursion concepts. For example, for sumArray, the MCQ asked about the role of the recursive step, and the OE asked to compare recursion to a loop-based approach. When both snippets were complete, they rated the explanations on two Likert items and provided two open-ended reflections on their usefulness. See Figure~\ref{fig:flow} for an overview of the study procedure. This design allowed some students to experience both explanation types across snippets.

\subsection{Data Analysis}
\subsubsection{Students' Performance}
To assess student comprehension, we analyzed student performance on both MCQ and OE questions. MCQ data were quantitatively evaluated by calculating the percentage of correct responses across each experimental condition (generic and diverse explanations), as well as comparing accuracy across different code snippets (sumArray, randomizeString, countChar). For OE questions, responses were qualitatively assessed through manual coding. A rubric was developed to categorize answers as correct or incorrect based on the accuracy and completeness of the conceptual explanations provided by students. Inter-rater reliability was assessed using Fleiss' Kappa~\cite{fleiss1971measuring}, which accounts for multiple raters. This resulted in a coefficient of $k=0.75$, indicating substantial agreement. After establishing reliability, remaining responses were divided among raters for independent coding. We used chi-square tests of independence to compare correctness rates across explanation conditions.
\subsubsection{Students' Preferences}

To approximate perceived cognitive load, we used two Likert-scale items measuring perceived helpfulness (1 = Strongly Disagree, 5 = Strongly Agree) and amount of information (1 = Not Enough to 5 = Too Much). These measures served as proxies for cognitive effort and perceived overload. For analysis, we calculated mean Likert scores per explanation group. We aggregated these scores without differentiating by snippet, providing overall comparative insights into students' perceived helpfulness and the perceived amount of information. Descriptive statistics and Kruskal-Wallis H tests were conducted to determine whether significant differences existed between explanation groups.

\subsubsection{Thematic Analysis}

To complement the performance analysis, we analyzed students' responses to the open-ended feedback questions using a thematic analysis based on Braun and Clarke's guidelines~\cite{braun_using_2006}. Responses were initially reviewed by four independent researchers who generated an initial set of open codes. These codes captured students' perceptions of helpfulness, clarity, cognitive load, and issues they encountered (e.g., redundancy, confusion). After independently reviewing 10\% of the responses, coders met to discuss codes, define thematic categories, and develop a final coding framework. The remaining responses were then independently coded with periodic cross-checks for consistency.

\section{Results}
\subsection{RQ1: Impact on Understanding}

We evaluated the impact of various LLM-generated explanations on student understanding by analyzing open-ended (OE) and MCQ assessments. In MCQ assessments (see Table~\ref{table:mcq_accuracy}), which tested students' grasp of programming concepts, accuracy was consistently high (>87\%) across all conditions, with negligible differences between diverse and generic explanations.
\begin{table}[htbp]
\centering
\caption{Student accuracy (percentage correct) on MCQs across conditions, including total responses per group.}
\label{table:mcq_accuracy}
\vspace{-10pt}
\begin{tabular}{l l c c c}
\textbf{Code Snippet} & \textbf{Type} & \textbf{Correct} & \textbf{Total} & \textbf{Accuracy} \\
\toprule
\textbf{sumArray}       & Diverse                   & 446              & 502            & 88.84                  \\
                    & Generic                   & 412              & 469            & 87.84                  \\
        \midrule

\textbf{randomizeString}     & Diverse                   & 238              & 244            & 90.16                 \\
                    & Generic                   & 219              & 235            & 95.32                  \\
        \midrule

\textbf{countChar}       & Diverse                   & 238              & 258            & 92.25                  \\
                    & Generic                   & 219              & 234            & 93.59   \\\bottomrule              
\end{tabular}
\vspace{2pt}
\par\noindent{\small\textit{Note: Chi-square tests found no significant differences} \\\textit{in MCQ accuracy between conditions ($p > .05$).}}

\end{table}

Both explanation types supported comprehension on MCQ items. In the OE assessments (see Table~\ref{tab:performance}), we observed a consistent improvement among students who received diverse explanations. A total of 785 students participated. The \textit{sumArray} group that received diverse explanations improved 7.74\% over the generic condition. The \textit{randomizeString} group exhibited an 8.1\% improvement, and the \textit{countChar} group showed a 7.7\% increase. We modeled OE correctness with logistic regression; the condition effect did not remain significant after Bonferroni adjustment (z = 2.15, p = 0.095).

\begin{table}[h]
    \centering
    \renewcommand{\arraystretch}{1.2}
    \caption{Performance comparison between Diverse and Generic categories, across three problems (P1, P2, and P3). }
    \label{tab:performance}
    \vspace{-10pt}
    \begin{tabular}{lccc}
        \toprule
        \textbf{Category} & \textbf{Total} & \textbf{Correct} & \textbf{\% Correct} \\
        \midrule
        \textit{sumArray} - Diverse  & 403 & 193 & 47.90 \\
        \textit{sumArray} - Generic & 381 & 153 & 40.16 \\
\textbf{Accuracy Increase} & & & \textbf{+7.74} \\        
        \midrule
        \textit{randomizeString} - Diverse  & 187 & 123 & 65.80 \\
        \textit{randomizeString} - Generic & 201 & 116 & 57.70 \\
        \textbf{Accuracy Increase}   &  &  & \textbf{+8.10} \\
        \midrule
        \textit{countChar} - Diverse   & 211 & 148 & 70.10 \\
        \textit{countChar} - Generic  & 186 & 116 & 62.40 \\
        \textbf{Accuracy Increase}   &  &  & \textbf{+7.70} \\
        \bottomrule
    \end{tabular}
    \vspace{2pt}
    \par\noindent{\small\textit{Note:} OE correctness modeled via logistic regression with Bonferroni adjustment; condition effect was not significant ($z = 2.15$, $p = .095$).}
\end{table}

\subsection{RQ2: Perceived Helpfulness and Usefulness}

RQ2 investigates how students perceived the helpfulness and amount of information provided by diverse versus generic explanations. We analyzed students' perceptions of explanation helpfulness through Likert-scale responses (\textit{1=Strongly Disagree, 5=Strongly Agree}) without differentiating by code snippet. The group receiving diverse explanations for both questions (DD) reported the highest perceived helpfulness ($Mean=3.68$). Students who experienced diverse explanations second (GD; $Mean=3.64$) closely followed. Those consistently receiving generic explanations (GG; $Mean=3.55$) and those transitioning from diverse to generic (DG; $Mean=3.50$) rated helpfulness slightly lower. A Kruskal-Wallis H test found no statistically significant differences in helpfulness ratings across groups, $H(7) = 4.42$, $p = .730$. This non-parametric test was used due to the ordinal nature of Likert-scale data.

We also assessed students' perceptions regarding the amount of information in the explanations using a Likert scale (\textit{1=Not Enough Information, 5=Too Much Information}). Students reported similar perceptions of the amount of information provided across all groups, with mean ratings around 3.3–3.4, indicating that students generally found the amount of information appropriate. The very minor differences between groups suggest that diverse explanations did not result in perceived information overload compared to generic explanations. A Kruskal-Wallis H test also showed no significant difference in perceived information quantity across groups, $H(7) = 3.57$, $p = .828$, showing that students generally rated the information load similarly regardless of condition.

\subsubsection{Students' Perceptions of Diverse Explanations}

\begin{figure}[ht]
    \centering
    \includegraphics[width=0.95\linewidth]{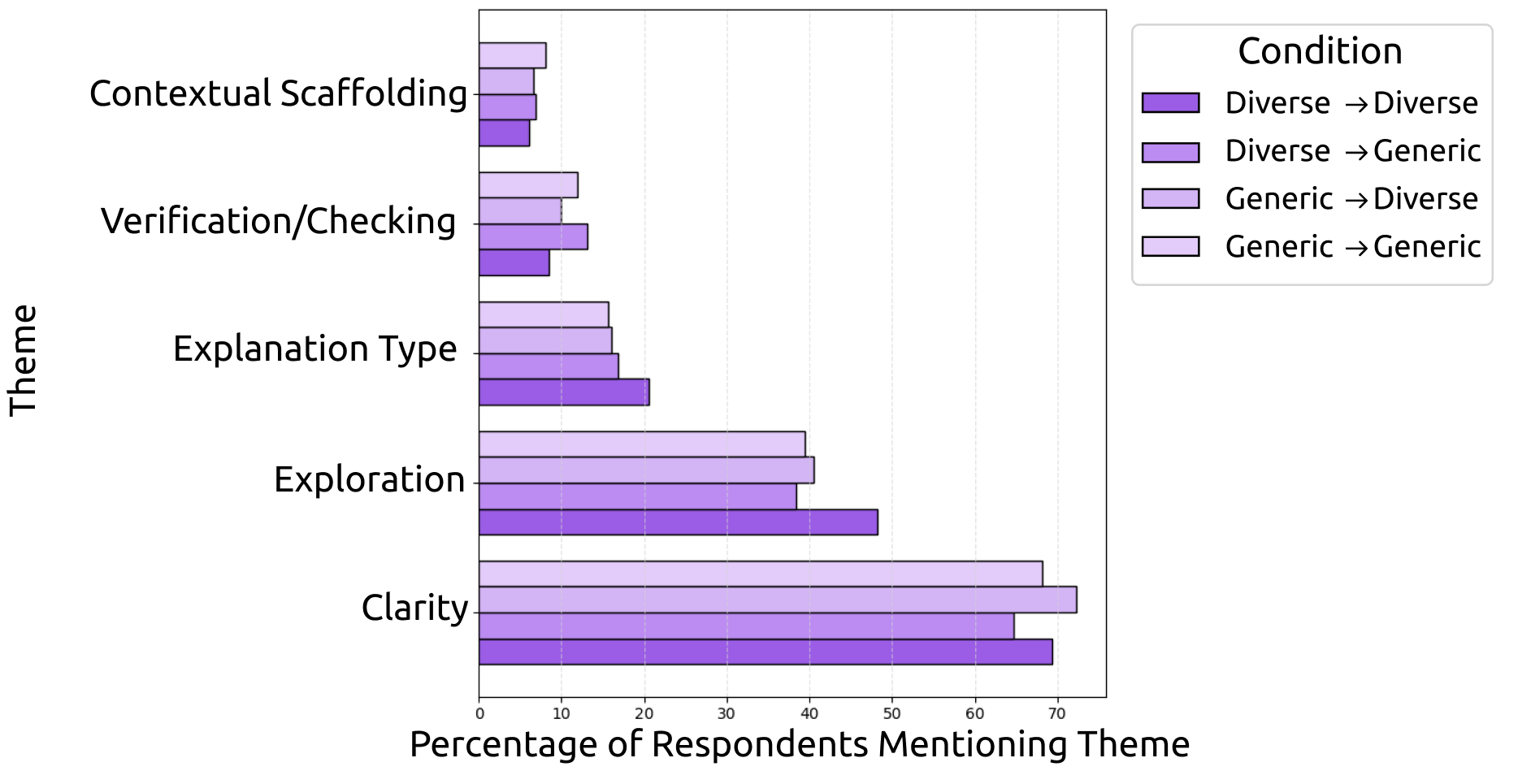}
    \caption{Most cited themes across explanation conditions for the question on what aspects students found most useful.}
    \label{fig:theme_bar_chart}
\end{figure}

In addition to analyzing performance, we examined students’ perceptions of the explanations using open-ended feedback. Students were asked, ``What aspect(s) of the explanations did you find most useful?'' Of the 785 students who provided OE responses, responses could reflect multiple themes. The most cited themes are shown in Figure ~\ref{fig:theme_bar_chart}, with lower-frequency themes (redundancy, confusion, too much information) grouped into the Overload/Frustration compound category in Table ~\ref{tab:compoundthemes}.

To better understand patterns in student reflections, we grouped commonly co-occurring themes into three higher-level categories:
\begin{itemize}
    \item \textbf{Engaged Understanding:} \\clarity OR exploration OR explanation type
    \item \textbf{Active Monitoring:} \\verification/checking AND clarity OR learning
    \item \textbf{Overload/Frustration:} \\too much information OR redundancy OR confusion
\end{itemize}


Engaged Understanding captures responses describing active meaning-making or conceptual clarity. Active Monitoring captures responses where students used explanations to verify or check their existing understanding. Overload/Frustration captures responses citing redundancy, confusion, or excessive information. These compound categories reflect different modes of cognitive engagement. Table~\ref{tab:compoundthemes} shows the percentage of students in each group whose responses reflected each compound theme.

To test whether these compound themes were associated with the explanation condition, we ran a chi-square test of independence for each cluster. While none reached statistical significance, residual analysis revealed consistent patterns. Students in the DD condition were more likely than expected to mention Engaged Understanding (resid = +0.48) and less likely to report Overload/Frustration (resid = $-$1.49). Conversely, students in the DG group were more likely to report Overload/Frustration (resid = +1.16).

\begin{table}[ht]
\centering
\caption{Compound themes for student perception.}
\label{tab:compoundthemes}
\vspace{-10pt}
\begin{tabular}{lccc}
\toprule
\textbf{Condition} & \textbf{Engaged} & \textbf{Monitoring} & \textbf{Overload} \\
\midrule
DD & 88.94\% & 8.54\% & 4.02\% \\
DG & 81.58\% & 10.53\% & 8.95\% \\
GD & 88.33\% & 9.44\% & 7.78\% \\
GG & 84.32\% & 7.57\% & 6.49\% \\
\bottomrule
\end{tabular}
\end{table}

Engaged Understanding was the most common theme across all conditions, especially among students who received diverse explanations last (DD, GD). This supports that diverse explanations facilitate deeper cognitive engagement as this was the most recent explanation type seen before answering. Active Monitoring, where students use explanations to verify their thinking was slightly more common in mixed conditions. Overload/Frustration appeared least often, but least in the condition with both explanations being diverse. These patterns are consistent with diverse explanations supporting deeper engagement, though the chi-square tests were not significant and the residuals alone do not support causal conclusions.

\subsubsection{Students' Use of Explanations}

In addition to the perception question, students were asked to ``Please reflect on and describe whether and how you used the explanations to aid your comprehension of the code.'' The most cited themes are shown in Figure~\ref{fig:theme_bar_chart_q12}. Their open-ended responses revealed different levels of engagement and were coded into multiple thematic categories. To capture broader patterns in explanation use, we clustered related themes into three higher-level categories:

\begin{figure}[ht]
    \centering
    \includegraphics[width=0.95\linewidth]{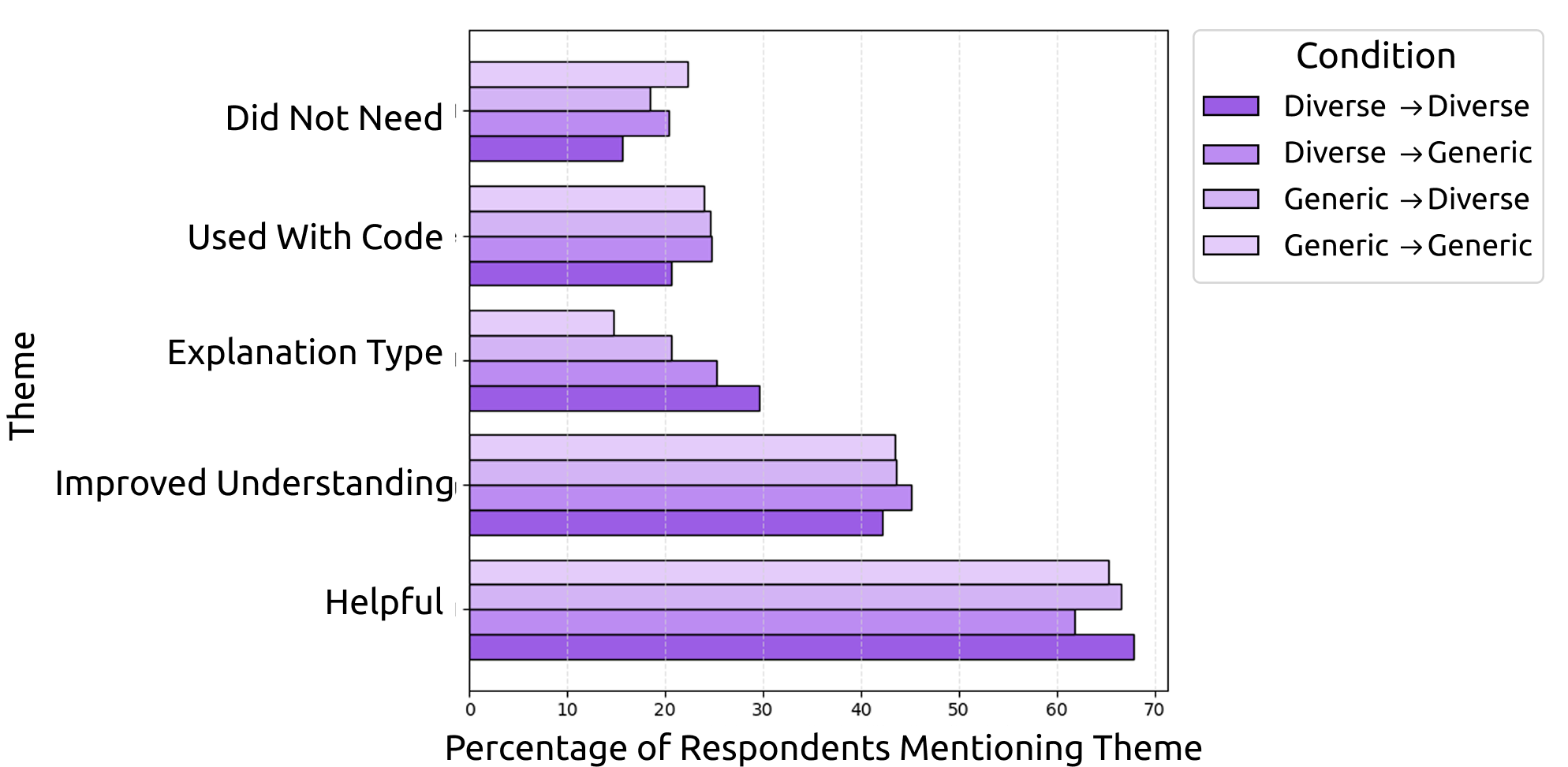}
    \caption{Most cited themes across explanation conditions for the question on how students used the explanations.}
    
    \label{fig:theme_bar_chart_q12}

\end{figure}

\begin{itemize}
    \item \textbf{Strategic Use:} Helpful, Used with code, Used different parts, Used own approach
    \item \textbf{Conceptual Support:} Improved Understanding, Explanation type helped
    \item \textbf{Dismissed:} Did not need, Did not read, Code too easy
\end{itemize}

No differences reached statistical significance (Table~\ref{tab:q12compoundthemes}), though residuals showed consistent patterns. DD students were less likely to dismiss explanations (resid = –1.21), while GG students were more likely to do so (resid = +0.72) and less likely to mention conceptual support (resid = –0.80).

\begin{table}[ht]
\centering
\caption{Compound themes for student uses.}
\vspace{-10pt}
\label{tab:q12compoundthemes}
\small
\begin{tabular}{lccc}
\toprule
\textbf{Condition} & \textbf{Strategic Use} & \textbf{Conceptual Support} & \textbf{Dismissed} \\
\midrule
DD & 77.89\% & 53.27\% & 16.08\% \\
DG & 75.27\% & 54.30\% & 20.97\% \\
GD & 81.01\% & 54.75\% & 20.67\% \\
GG & 76.09\% & 48.37\% & 22.28\% \\
\bottomrule
\end{tabular}
\end{table}

Strategic Use was the most frequently mentioned theme across all conditions, with over 75\% of students in each group describing the explanations as helpful or applicable while reading code. Conceptual Support was also common, particularly among students exposed to diverse explanations. Dismissive perspectives were least frequent overall, and lower among students who received diverse explanations in both stages.

\section{Discussion}
\subsection{Interpretation}

Our results did not show significant differences in performance between students who received diverse explanations and those who received generic ones. However, Variation Theory~\cite{lo_towards_2012} predicts that learning is most effective when learners are exposed to key variations across examples, which allows them to discern critical features and conceptual distinctions. While our intervention aimed to introduce diversity in the explanations provided, it is possible that the sources of variation that we employed were not aligned with the dimensions that students needed to discern in order to develop their understanding.


Cognitive Load Theory distinguishes between intrinsic load, driven by task complexity, and extraneous load, driven by avoidable processing demands~\cite{sweller_cognitive_2011,duran2021cognitive}. By presenting concise explanations, each focusing on a distinct dimension, the design likely reduced extraneous load, in turn improving understanding without increasing overall cognitive load~\cite{margulieux2012subgoal}. This design aligns with prior studies suggesting segmented instructional material is more cognitively manageable and effective~\cite{mayer_nine_2003}. Our results suggest, however, that reducing extraneous load alone may be insufficient if students do not engage deeply with the explanations. Explanation length and information density may shape how students interact with LLM-generated instructional content, and as generative AI makes explanations increasingly cheap to produce, ensuring student engagement with them becomes a more pressing design challenge.

\subsection{Instructional and Practical Implications}

Although the diverse explanations in this study did not provide substantial additional benefits to students, it is possible that students may need explanations that vary in more personally relevant ways. One potential direction is to explore relevance-driven diversity, where the variation among explanations aligns with learners’ individual interests or prior knowledge domains (for example, using design metaphors or pop-culture contexts). Such alignment also supports equity, as varied cultural lenses may help students who feel alienated by traditional examples better connect with the material. Recent work showed that students are more engaged by creating their own analogies with large language models~\cite{bernstein_like_2024,bernstein_analyzing_2024} and that they enjoy controlling the topics of exercises~\cite{logacheva2024evaluating}. 

These findings have practical implications for instructors and developers of automated tutoring systems. Rather than presenting a generic explanation repeatedly, LLMs could generate multiple, concise perspectives allowing students to select the angle they find most helpful, building on evidence that explanations are among the most common ways students use generative AI tools~\cite{hou2024effects}. This approach can support differentiated instruction at scale, personalized to diverse learning preferences. Further research could investigate the long-term retention effects of exposure to diverse explanations and explore whether optimal numbers or types of explanations vary with student expertise levels.

\subsection{Limitations}
While this study explores patterns associated with exposure to diverse LLM-generated explanations, several factors limit what we can conclude about their impact on students’ understanding of recursion. First, all participants came from a single course taught by the same instructor, and we did not measure prior knowledge of recursion. Second, students only saw two short programming exercises and were tested immediately, limiting the magnitude of any observable effects and potentially capturing short-term rather than durable learning. Finally, high overall scores point to a possible ceiling effect; when students already perform well, small improvements are harder to detect. We also cannot verify whether or how thoroughly students engaged with the explanations, and we did not compare outcomes across the three explanation dimensions individually, which limits conclusions about their instructional use. Students had no opportunity to ask follow-up questions or interact with the explanations dynamically, more closely resembling an instructor distributing curated written explanations than a typical LLM deployment. Future work should test broader tasks, repeated exposure, and interactive explanation systems to better measure the impact of explanation diversity.

\section{Conclusion}
While diverse explanations did not produce statistically significant gains in comprehension, they were associated with consistently higher performance and more engaged student responses across multiple measures. This suggests they might be a promising strategy for educators aiming to deepen learning through varied instructional content. Future research should explore the ideal amount and types of explanations and examine long-term retention and transfer of programming knowledge.

\begin{acks}
We thank Hannah V. Nguyen for creating the study flow diagram.
\end{acks}



\balance
\bibliographystyle{ACM-Reference-Format}
\bibliography{references}


\end{document}